\documentclass[%
 reprint,
 amsmath,amssymb,
 aps,
prb,
]{revtex4-1}

\usepackage[utf8]{inputenc}
\usepackage{graphicx}
\usepackage{dcolumn}
\usepackage{bm}
\usepackage{dcolumn}
\usepackage{dsfont}
\usepackage{bbm}
\usepackage{hyperref}
\usepackage[mathlines]{lineno}
\usepackage{natbib}

\usepackage[table]{xcolor}
\usepackage{amssymb}
\usepackage{amsmath}
\usepackage[utf8]{inputenc}
\usepackage[english]{babel}
\usepackage{graphicx}
\usepackage{dcolumn}
\usepackage{float}
\usepackage{bm}
\usepackage{blkarray}
\usepackage{multirow}
\usepackage[mathlines]{lineno}

\begin{document}

\title{Radiative Heat Transfer Between Core-Shell Nanoparticles}

\author{Moladad Nikbakht}
\email{mnik@znu.ac.ir}
\affiliation{%
 Department of Physics, University of Zanjan, Zanjan, Iran
}%

\date{\today}

\begin{abstract}
Radiative heat transfer in systems with core-shell nanoparticles may exhibit not only a combination of disparate physical properties of its components but also further enhanced properties that arise from the synergistic properties of the core and shell components. We study the thermal conductance between two core-shell nanoparticles (CSNPs). The contribution of electric and magnetic dipole moments to the thermal conductance depend sensitively on the core and shell materials, and adjustable by core size and shell thickness. We predict that the radiative heat transfer in a dimer of Au@SiO$_2$ CSNPs (i.e., silica-coated gold nanoparticles) could be enhanced several order of magnitude compared to bare Au nanoparticles. However, the reduction of several orders of magnitude in the heat transfer is possible between SiO$_2$@Au CSNPs (i.e., silica as a core and gold as a shell) than that of uncoated SiO$_2$ nanoparticles.
\end{abstract}


\maketitle

\section{Introduction}\label{sec1}

Radiative heat transfer between objects is sensitively dependent upon their separation distance. If the separation is too small compared to the thermal wavelength, then energy transfer exceeds the well-known classical Planck’s law of the black-body radiation \cite{hu}. Several studies have shown that the radiative heat transfer between objects with planar geometry depends on the materials and could be manipulated by using anisotropic materials, layered materials, and covering objects with different material composition \cite{Hyperbolic,nonlocal,DIDARI2018120,covered,chang,BAI201536}. In addition to the separation distance, the radiative heat transfer in a dimer of nanoparticles depends on various properties of the constituent nanoparticles, including material composition \cite{chapuis,manybodyplasmonic}, size \cite{threebodysizeeffect}, surface structure \cite{spin-like}, shape \cite{Ramezan,shapeefect,shap}, and relative orientation \cite{Nikbakht,nikbakht1,threebodysplitter}. From these studies, it can be concluded that, in addition to the dielectric constant of the host material, the radiative heat transfer strongly depends on the optical properties of objects in the system. Since the optical properties of particles strongly depend on their characteristics, the shape, size, and material composition of particles have a decisive role in heat transfer, which have been the subjects of majority of studies in past years. The proper choice of each of these quantities depends on the desired application which could include reducing, increasing or rectifying the radiative heat transfer. 

Material selection is of great importance in thermal management and among materials, those that can support surface phonon modes in the infrared (e.g., polar dielectrics SiC and SiO$_2$) are generally used to enhance near-field heat transfer. On the contrary, metallic nanoparticles which have resonances only in the visible or the UV range, do not perform as well as polar dielectrics for near-field heat transfer applications. The reason is the negligible contribution of plasmon modes to thermal transport, since according to Wien displacement law, the peak wavelength for thermal emission at room temperature, is much longer than the surface plasmon wavelengths of most conventional metals. However, the search for methods that can causes a red-shift in the surface plasmon resonance of metals, can allow the ability to tailor the spectrum of radiative heat transfer to achieve an enhancement in radiative transfer between metal nanoparticles. Pioneer work by Pendry {\it et al}.\cite{Pendry} demonstrated that structured surfaces can be used to  engineer  a  surface  plasmon  on  metals at  almost  any  frequency. Recently, Braden {\it et al}.\cite{heattransfercoated} have shown that coated spheres with silver cores and silica coatings have different spectrum of radiative heat transfer than that of homogeneous spheres of either of its constitutive components. Similarly, the plasmon hybridization picture can be used to tune the surface plasmon modes of spheres coated with a metallic layer. 

\begin{figure}[t]
\centering
\includegraphics[height=4cm,width=8.3cm]{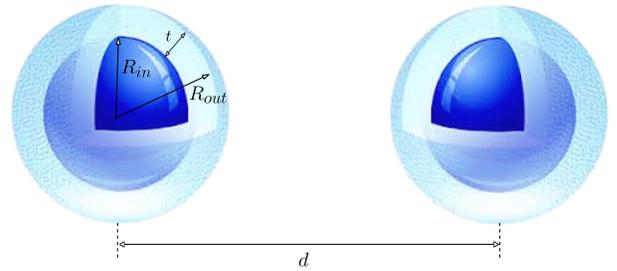}
\caption{Scheme of the system of CSNPs under study. System consists of two identical core-shell nanoparticles with separation $d$. Each nanoparticle characterized by its core radius $R_{in}$, outer radius $R_{out}$, shell thickness $t=R_{out}-R_{in}$, core volume fraction $f=(R_{in}/R_{out})^3$, permittivity of the inner layer $\epsilon_{c}$, and permittivity of the outer layer $\epsilon_{s}$.}
\label{fig1}
\end{figure}
This is an exciting time for solar thermal collectors and energy management applications to manipulate the thermal flux by nanoparticles \cite{fluid1,ALGEBORY2018625}. For such applications, novel ways for tuning the rate of energy transfer is important, and a unique method for this idea is through the use of core-shell nanoparticles (CSNPs). Core-shell nanoparticles have received increasing attention since Halas and co-workers \cite{firsthollow}. During the last decade, a variety of approaches have been developed for the synthesis of CSNPs, which provide us with metallic (or dielectric) nanoparticles coated of a shell with a different composition (e.g., Au on SiO$_2$ or SiO$_2$ on Au). Several approaches to fabrication of CSNPs based on the silica core and gold shell or gold core and silica shell have been reported\cite{ausio21,ausio2synthesis,sio2au2,sio2au3,sio2ausynthesis1,sio2ausynthesis2,ausio2synthesis2,ausio2synthesis3}. It is shown by several groups that CSNPs exhibit unique optical properties which cannot be obtained by traditional nanoparticles with uniform structure\cite{coreshellreview1,coreshellreview2,ausio2synthesis4optic}. With the use of core-shell spherical nanoparticles, one has two dielectric functions, the core radius and the shell thickness as parameters, instead of the single dielectric function and size as tuning parameters for a bare nanoparticle.

An example that can illustrate the significance of the study of radiative transfer between CSNPs is the emulsion of metal nanoparticles in an aqueous solution. Due to the chemical interaction at the surface of the nanoparticles, interface between nanoparticles and solution can be viewed as a double layer, suggesting that such nanoparticles can be modeled by CSNPs. Hence, the radiative heat transfer in such systems can be modeled with core@shell nanoparticles. Moreover, growing interest of researchers in the measuring thermal near fields has led to new techniques, such as quantitative measurements of the near-field heat flux between the scanning thermal microscope tip and a planar sample at nanometer distances. An important aspect of this topic is the fact that, for theoretical predictions, the tip is usually modeled as sphere, however, in many of these experiments a coated tip is used in
measurement of near-field thermal properties in probe-sample setups. It is also shown that insufficiently cleaned probe (which acts as a coating layer) or coverage material in such probe-sample experiments leads to drastic change in the thermal conductance \cite{tipsampelexp}. Accordingly, the unique properties of coated particles should be studied and is important, because it provides insight into the advantageous use of CSNPs in thermal managements.

In this paper, we investigate the radiative transfer between core-shell nanoparticles. We show that using CSNPs can drastically change the radiative heat transfer compared with the radiative transfer between single component nanoparticles, suggesting the presence of new modes participating in the heat flux between such nanoparticles. There are two types of CSNP investigated in this research, i.e. SiO$_2$@Au and Au@SiO$_2$ core-shell nanoparticles. We demonstrate that the radiative heat transfer in a dimer of Au@SiO$_2$ (or SiO$_2$@Au) CSNPs not only combine the optical signatures of silica and the plasmonic properties of gold, but exhibits further enhanced and expanded plasmonic tunability as well due to the dielectric properties of the shell material surrounding the core. In addition, the electric and magnetic dipole contributions to heat transfer depend on the core and shell materials, and changing the size of the core or the thickness of the shell leads to a change in their contribution in the heat transfer. The combination of these effects could result in a giant enhancement of the thermal conductance in a dimer of silica-coated Au nanoparticles compared to that of bare Au nanoparticles. On the other side, we show that reduction of several order of magnitude in the heat transfer is possible between gold-coated silica nanoparticles (SiO$_2$@Au CSNPs) than that of uncoated SiO$_2$ nanoparticles.

The structure of the paper is as follow. The formalism is developed in Sec.~\ref{sec2}, where thermal conductance in a dimer of CSNPs is derived in terms of fluctuating electric and magnetic diple moments. In Sec.~\ref{sec3} we have studied the thermal properties of metal nanoparticles coated with a dielectric layer. We have used Au@SiO$_2$ as a typical CSNP, and the influence of SiO$_2$ thickness and Au volume fraction on the thermal conductance, are discussed. A similar study was carried out on the heat transfer between SiO$_2$ nanoparticles (as a dielectric core), coated with gold layer (as a metal shell), in Sec.~\ref{sec4}. Finally, our work is summarized in Sec.~\ref{sec5}.

\section{Model}\label{sec2}
The schematic of the system under consideration is sketched in Fig.~(\ref{fig1}). Two spherical CSNPs are located at fixed distance of $d$ from each other inside a transparent dielectric of permittivity
$\epsilon_h$. Nanoparticles are kept at temperatures $T_i$ ($i=1,2$), and exchanges energy via radiation. Each nanoparticle is characterized by its core radius and shell thickness denoted by $R_{in}$, and $t=R_{out}-R_{in}$ respectively; where $R_{out}$ is the radius the nanoparticle. Moreover, the core volume fraction $f=(R_{in}/R_{out})^3=(1+t/R_{in})^{-3}$ is introduced, which represents the ratio of the core volume to the total volume of each CSNP. We assume that the size of nanoparticles are small compared to the thermal wavelength $\lambda_{T}=c\hbar/k_{B}T$ (wherein $c$ is light velocity in the vacuum, $2\pi\hbar$ and $k_{B}$ are Planck and Boltzmann constants, respectively), and nanoparticles exchange energy due to dipolar interaction (for more information on the radiative transfer between large CSNPs see Ref.~\cite{heattransfercoated}). With this assumption and using the Mie's first coefficients \cite{bohrenbook}, we approximated each CSNP by fluctuating dipole with electric and magnetic polarizabilities:

\begin{subequations}
\label{eq1}
\begin{eqnarray}
\label{eq1a}\alpha^E&=&3v\frac{(\epsilon_s-\epsilon_h)(\epsilon_c+2\epsilon_s)+f(\epsilon_c-\epsilon_s)(\epsilon_h+2\epsilon_s)}{(\epsilon_s+2\epsilon_h)(\epsilon_c+2\epsilon_s)+f(\epsilon_c-\epsilon_s)(2\epsilon_s-2\epsilon_h)},~~~~~\\
\label{eq1b}\alpha^H&=&\frac{v}{10}(k_hR_{out})^2\Big[\frac{(\epsilon_s-\epsilon_h)+f^{5/3}(\epsilon_c-\epsilon_s)}{\epsilon_h}\Big],
\end{eqnarray}
\end{subequations}
where $v=\frac{4}{3}\pi R_{out}^3$ is the volume of the CSNP,~$k_h=\sqrt{\epsilon_h}\omega/c$,~and $\epsilon_c(\omega)$ and $\epsilon_s(\omega)$ are the dielectric permittivity of core and shell, respectively. It is easy to show that, for $\epsilon_c=\epsilon_s$, the polarizanilities in Eq.~(\ref{eq1}) reduce to those for a homogeneous sphere. 
To simplify our analysis, we assume the radiative heat transfer in dimer with identical CSNPs. To this end, CSNPs are assumed to have same core (i.e., $\epsilon_{c}\equiv\epsilon_{1,c}=\epsilon_{2,c}$ , $R_{in}=R_{1,in}=R_{2,in}$) and also same shell (i.e., $\epsilon_{s}\equiv\epsilon_{1,s}=\epsilon_{2,s}$, $t=t_{1}=t_{2}$). Moreover, we take the dielectric function $\epsilon_h$ of the surrounding medium to be 1. The heat transfer between two CSNP is modeled by the interaction of simple fluctuating dipoles and the coupled electric-electric and magnetic-magnetic dipole approach \cite{manybodyzhao} is used to calculate the contribution of electric and magnetic dipole moments to the radiative heat flux. The associated net heat flux can be described using the many-body radiative heat transfer theory \cite{generalformalism,nikbakht1,ben1,manybodyzhao}, from which the mutual conductance at small temperature mismatch ($T_1-T_2\rightarrow 0$) around temperature $T$ reads \cite{generalformalism}
\begin{equation}
\mathcal{G}=\sum_{\nu=E,H}\int_{0}^{\infty} \frac{d\omega}{2\pi} \mathcal{T}_\nu(\omega)\frac{\partial{\Theta(\omega,T)}}{\partial{T}}, 
\label{eq2}
\end{equation}
where $\Theta(\omega,T)=\hbar\omega/[e^\frac{\hbar\omega}{K_{B}T}-1]$ is the mean energy of a Plank harmonic oscillator in thermal equilibrium at frequency $\omega$ and temperature $T$. The transmission coefficient, $\mathcal{T}_\nu(\omega)$, with contribution of electric ($\nu=E$) and magnetic ($\nu=H$) dipole moments, denotes the energy transmission probability between these nanoparticles at frequency $\omega$, and could be defined as \cite{Ramezan}
\begin{equation}
\mathcal{T}_\nu(\omega)=2\mathrm{Im}(\chi_1^\nu)\mathrm{Im}(\chi_2^\nu)\mathrm{Tr}\Bigg[\frac{{\hat{\mathrm G}}
{\hat{\mathrm G}}^\dagger}{\hat{\mathrm M}_\nu{\hat{\mathrm M}}_\nu^\dagger}
\Bigg]~~~(\nu=E,H).
\label{eq3}
\end{equation}
Here $\chi^\nu=\alpha^\nu+\mathrm{G}_{0}^*|\alpha^\nu|^2$ is the susceptibility of CSNP with polarizability $\alpha^\nu$, and $\mathrm{G}_{0}=i(k^3/6\pi)$. Moreover, $\hat{\mathrm{M}}_\nu=[\mathbbm{1}-\alpha^\nu_1\alpha^\nu_2\hat{\mathrm{G}}^2]$ accounts for the scattering of radiation fields between the two CSNPs, where $\mathbbm{1}$ stands for identity operator. The interaction of nanoparticles in a dimer is accounted for by the free space Green tensor:
\begin{equation}
\hat{\mathrm{G}}=\frac{k^3e^{ikd}}{4\pi}\Big[(\frac{k^2d^2+ikd-1}{k^3d^3})\mathbbm{1}+\frac{3-3ikd-k^2d^2}{k^3d^3}\frac{\bm{d}\otimes\bm{d}}{d^2}\Big],~~~~~~~~
\label{eq4}
\end{equation}
where $\bm{d}$ is the vector linking the center of two CSNPs. The thermal conductance due to electric polarization (known as {\it electric-electric} or "EE" interaction) has been extensively studied in the past, and its magnetic counterpart (known as {\it magnetic-magnetic} or "MM" interaction) is generally regarded as a weak contribution. However, there are situations in which the magnetic contribution to heat transfer can even exceed the electric contribution\cite{chapuis}, specifically in common metallic nanoparticles (with large sizes) exchanging energy in the far-infrared range. Moreover, as shown by Dong {\it et al}. \cite{manybodyzhao}, Chen {\it et al}. \cite{CHENcoreshell} and Manjavacas {\it et al}. \cite{heatbetweenmetall2}, apart from the electric-electric and magnetic-magnetic interactions, cross terms of electric and magnetic response (known as {\it electric-magnetic} or "EM" term, and {\it magnetic-electric} or "ME" term) also contribute to the radiative heat flux. They have shown that each of the four terms can dominate the radiative heat transfer depending on the position and composition of particles. These studies demonstrate that the radiative heat transfer in a dimer of homogeneous polar (metal) nanoparticles is dominated by EE (MM) term. However, electromagnetic crossed terms (EM and ME) should not be ignored, since they could play a leading role in heat transfers within heterogeneous structures. Utilization of CSNPs, provide another freedom of tunability for controlling the contribution of EE, MM, EM and ME terms to heat transfer. It should be noted that, in the case of near field thermal conductance in a dimer of identical CSNPs used in our study, the structure is homogeneous and the EM and ME terms are quantitatively the same. On the other hand, since the thermal conductance in such structures is dominated by either EE or MM terms, the radiative transfer due (EE+MM) term largely exceeds the (EM+ME=$2\times$EM) term. Hence, we have omitted the two cross terms in our calculations and only take into account the contribution of EE and MM terms in thermal conductance. 

Apart from the configuration resonances\cite{generalformalism}, the transmission probability between CSNPs can show resonance (enhancement) due to the enhancement in $\mathrm{Im}(\chi^E)$ terms appeared in Eq.~(\ref{eq3}). The resonance condition for the electric dipole moment of a core@shell sphere, follows by setting the denominator of Eq.~(\ref{eq1a}) equal to zero. Accordingly, the resonances frequencies, which correspond to localized surface modes, are solutions of the following transcendental equation
\begin{equation}(\epsilon_s+2)(\epsilon_c+2\epsilon_s)+f(\epsilon_c-\epsilon_s)(2\epsilon_s-2)=0,\label{eq5}\end{equation}
From Eq.~(\ref{eq1a}), in the limit of zero core radius (i.e., $f=0$) or zero shell thikness (i.e., $t=0\Rightarrow f=1$), the resonance condition reduces to those for a homogeneous sphere occurring at $\epsilon_s=-2$ or $\epsilon_c=-2$, respectively. Depending on the material composition of core or shell, this is the resonance condition for the excitation of the surface plasmon (resp. phonon-polariton) modes of individual metal (resp. polar) nanoparticles, mainly known as the lowest-order surface mode. This is the dominant mode participating in the heat transfer in a dimer of nanoparticle. 
Equation~(\ref{eq5}) can be expressed in more compact form as
\begin{equation}(\epsilon_s-\varepsilon^+)(\epsilon_s-\varepsilon^-)=0\label{eq6}\end{equation}

where $\varepsilon^-$ and $\varepsilon^+$ are the permittivity value for the symmetric and antisymmetric
resonance modes, respectively. The excitation of these modes follows by setting $\epsilon_s(\omega)=\varepsilon^\pm$.
If we take the dielectric function of the core mediums ($\epsilon_c$) to be 1, we would have a hollow nanoparticle. From Eq.~(\ref{eq1a}), the electric polarizability of a hollow nanoparticle takes the form
\begin{equation}
\alpha^E=3v\frac{(\epsilon_s-1)(1+2\epsilon_s)(1-f)}{(\epsilon_s-\varepsilon^+)(\epsilon_s-\varepsilon^-)},
\label{eq7}
\end{equation}
resulting in the following resonant conditions \cite{bohrenbook}
\begin{equation}
\varepsilon^\pm=[-(5+4f)\pm3\sqrt{1+8f}]/(4-4f).
\label{eq8}
\end{equation}
In a limit of thick shell, i.e., $f\rightarrow 0$, we have $\varepsilon^+=-\frac{1}{2}+\frac{3}{2}f+\mathcal O (f^2)$ which corresponds to the resonance condition for the empty spherical volume surrounded by a medium with dielectric function $\epsilon_s$. As discussed earlier, the resonance of this mode vanishes for the limiting case $f=0$. In addition, another resonance occurs at $\varepsilon^-=-2-6f+\mathcal O (f^3)$ in this limit, which is the resonance condition for the excitation of symmetric modes. On the other side, we note that in the limit of an extremely hollow nanoparticle, i.e., $f\rightarrow 1$ $\Rightarrow$ $t\rightarrow 0$, we have $\varepsilon^+=-\frac{2}{3}f_c^{1/3}+\mathcal O (f_c^{2/3})$ and $\varepsilon^-=\frac{1}{2}-\frac{3}{2}f_c^{-1/3}-\frac{1}{3}f_c^{1/3}+\mathcal O (f_c^{2/3})$ with $f_c=1-f$. We observe that, resonance modes are functions of the hollow nanoparticle core volume fraction. As the volume fraction increases from 0 to 1, the resonance of the antisymmetric mode rises from -0.5 to 0, while the resonance of the symmetric mode is reduced from -2 to $-\infty$. 

In the case of CSNPs, the material composition of the core might have a large impact on the position of the resonance of dipolar excitations. In a limit of thick shell, the resonant modes can be described by two distinctive cases \cite{coreshellresonance}
\begin{equation}
\label{eq9}
\varepsilon^+=-\frac{\epsilon_c}{2}-\frac{3\epsilon_c}{2}\frac{\epsilon_c+2}{\epsilon_c-4}f~~~,~~~\varepsilon^-=-2+6\frac{\epsilon_c+2}{\epsilon_c-4}f~~~~~~{\tt for}~~~~\epsilon_c<4
\end{equation}
and
\begin{equation}
\label{eq10}
\varepsilon^+=-2+6\frac{\epsilon_c+2}{\epsilon_c-4}f~~~,~~~\varepsilon^-=-\frac{\epsilon_c}{2}-\frac{3\epsilon_c}{2}\frac{\epsilon_c+2}{\epsilon_c-4}f~~~~~~{\tt for}~~~~\epsilon_c>4
\end{equation}
We observe that the two resonant modes mutually flip their attribution beyond the critical value of $\epsilon_c=4$, and approach the plasmonic resonant condition of a solid sphere for $f=0$ . Similar behavior of this property holds for the resonant modes of thin-layer sphere beyond the critical value of $\epsilon_c=-2$, viz.,
\begin{equation}
\label{eq11}
\varepsilon^+=-2\frac{\epsilon_c}{\epsilon_c+2}f_c^{1/3}~~~,~~~\varepsilon^-=\frac{\epsilon_c}{2}-\frac{\epsilon_c+2}{2}f_c^{-1/3}~~~~~~{\tt for}~~~~\epsilon_c>-2
\end{equation}
and 
\begin{equation}
\label{eq12}
\varepsilon^+=\frac{\epsilon_c}{2}-\frac{\epsilon_c+2}{2}f_c^{-1/3}~~~,~~~\varepsilon^-=-2\frac{\epsilon_c}{\epsilon_c+2}f_c^{1/3}~~~~~~{\tt for}~~~~\epsilon_c<-2
\end{equation}

Regarding the shape of the dielectric function of the shell material, the dielectric function of the core layer controls the position of the resonance modes. In case of CSNPs with polar shell material, e.g., Au@SiO$_2$, the condition for the resonance of surface phonon modes could be satisfied only in a small frequency range where $\epsilon'_{SiO_2}(\omega)$ is negative. With these interpretations, it is expected that in a dimer of Au@SiO$_2$ CSNPs, a change in core volume fraction does not result in a dramatic change in the resonance of surface phonon modes. In contrast, for metal coated nanoparticles, e.g., SiO$_2$@Au, the frequency region in which $\epsilon'_{Au}(\omega)$ can take negative values is not restricted to a narrow bound. Therefore, the resonance of surface plasmon modes of metal coated nanoparticles could be tuned over the wide range of frequencies by changing the core volume fraction. Moreover, the resonance of symmetric modes exhibits a plasmonic resonance at lower energies than the antisymmetric case and the permittivity value for these modes diverge as $f$ increase. By appropriate choice of a core volume fraction, one can increase the participation of surface plasmon modes in the heat flux between metal coated nanoparticles even at temperatures around $T\approx300$~K, which would be of great interest in plasmonic thermal transport applications.

Let us now apply this theoretical framework to describe the thermal conductance in a dimer of CSNPs. As we mentioned earlier, two types of CSNPs (SiO$_2$@Au and
Au@SiO$_2$) are investigated in this research. We focus on the analysis of the thermal conductance at room temperature (300~K). To determine if a resonance in the transmission probability is realizable, where it occurs, its strength, and its contribution to heat transfer; we need to know how the core/shell dielectric functions vary with frequency. We have used the Drude-Lorentz model for the dielectric function of SiO$_2$ part of the effective polarizabilities of CSNPs\cite{palik1991handbook}. In order to make better link with experiment, we have used Drude critical point model for the dielectric function of bulk Au \cite{audielectricbulk}. Moreover, this dielectric function is modified in order to take into account the finite size effects in small core radius or small shell thicknesses. We rely on the following radius-dependent dielectric function for the metallic part of the CSNPs \cite{epsiloncorrection}
\begin{equation}
\epsilon_{Au}(\omega)=\epsilon_{bulk}(\omega)+\frac{\omega_p^2}{\omega^2+i\gamma_{bulk}\omega}-\frac{\omega_p^2}{\omega^2+i\gamma\omega},
\label{eq6}
\end{equation}

where $\omega_p=8.9234$~eV, $\gamma_{bulk}=0.042389$~eV, and $v_f=1.4\times10^6$~m/s. Moreover $\gamma=\gamma_{bulk}+v_f/L_{eff}$ is a radius-corrected relaxation and $v_f$ is the Fermi velocity of conduction electrons. There are different approximate formulas in the literature for effective mean free path of electrons ($L_{eff}$) which depend on the underlying model of electron scattering from the metal boundaries \cite{sizecorrections,shellcorrection}. We assume that $L_{eff}=\frac{4}{3}R_{in}$ for Au@SiO$_2$ CSNPs in which gold is a core material and $L_{eff}=\frac{4}{3}\frac{R_{out}^3-R_{in}^3}{R_{out}^2+R_{in}^2}$ in case of SiO$_2$@Au CSNPs in which gold is a shell material.

\section{Thermal conductance in a dimer of Au@SiO$_2$ core-shell nanoparticles}\label{sec3}

\begin{figure}[t]
\centering
\includegraphics[height=5.25cm,width=7.75cm]{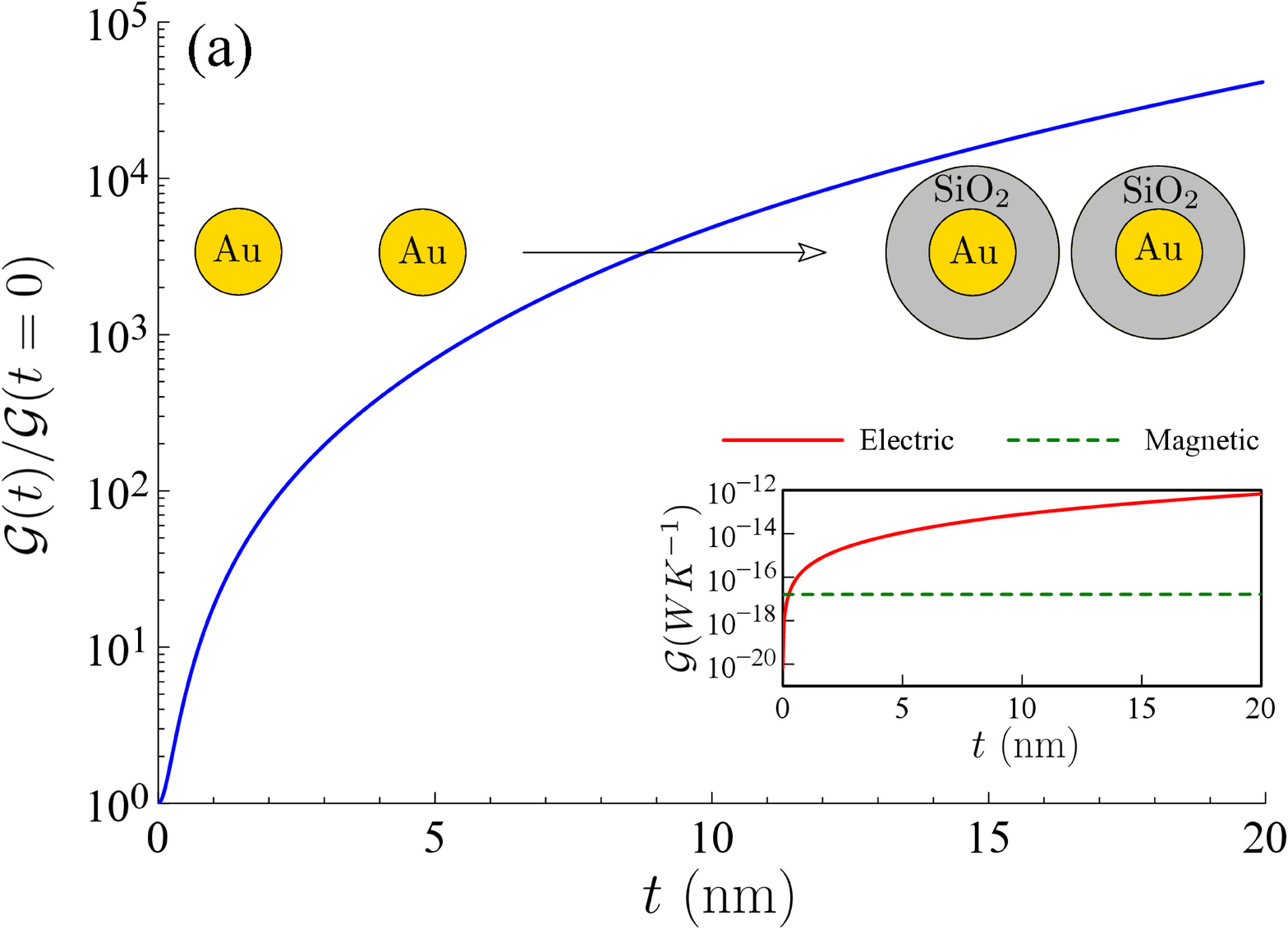}
\includegraphics[height=5.25cm,width=7.35cm]{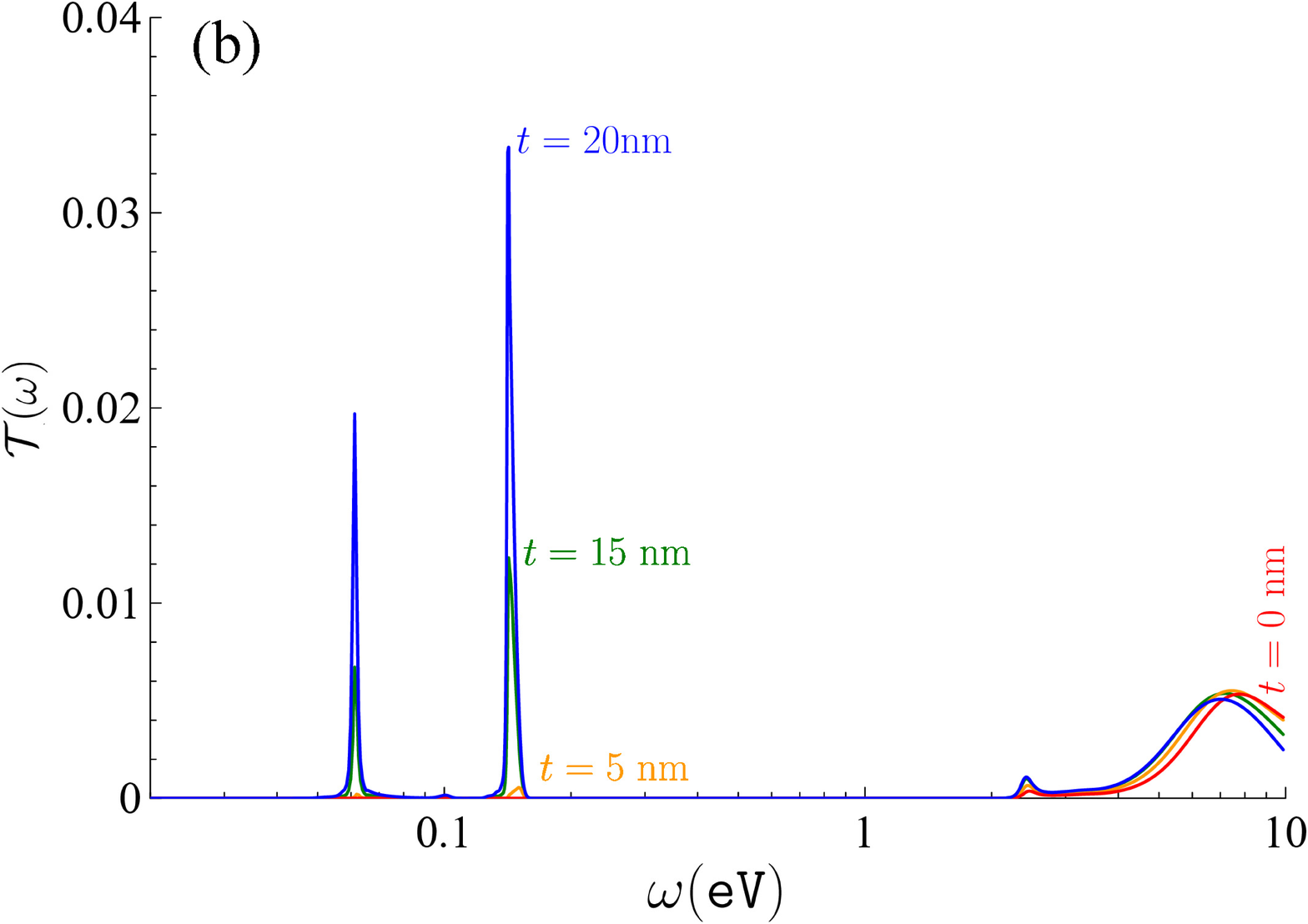}
\includegraphics[height=5.25cm,width=7.35cm]{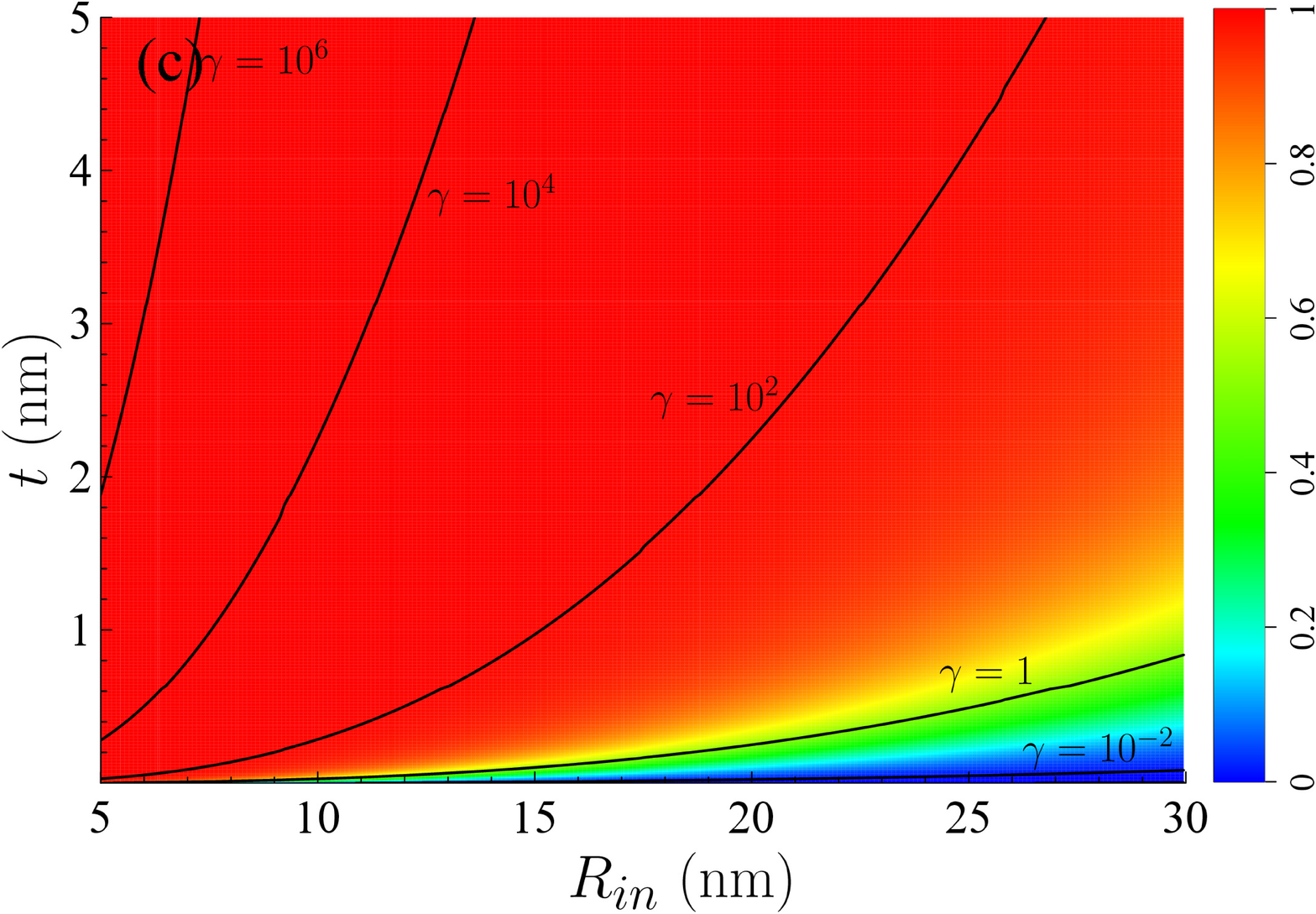}
\caption{(a) Normalized thermal conductance in a dimer of identical Au@SiO$_2$ CSNPs with $d=120$~nm for fixed gold radius ($R_{in}=20$~nm) as a function of the shell (silica) thickness. The inset graph shows the contribution of the electric ($\mathcal{G}_E$) and magnetic ($\mathcal{G}_H$) dipole moments to the thermal conductance. (b) The transmission probability of the system in part a. (c) Relative contribution of electric dipole to the thermal conductance ($\mathcal{G}_E/(\mathcal{G}_E+\mathcal{G}_H)$) and the ratio of the electric to magnetic dipole contributions ($\gamma=\mathcal{G}_E/\mathcal{G}_H$) in a dimer of identical Au@SiO$_2$ CSNPs as a function of Au radius and SiO$_2$ thickness. }
\label{fig2}
\end{figure}

\begin{figure}[t]
\centering
\includegraphics[height=5.25cm,width=7.75cm]{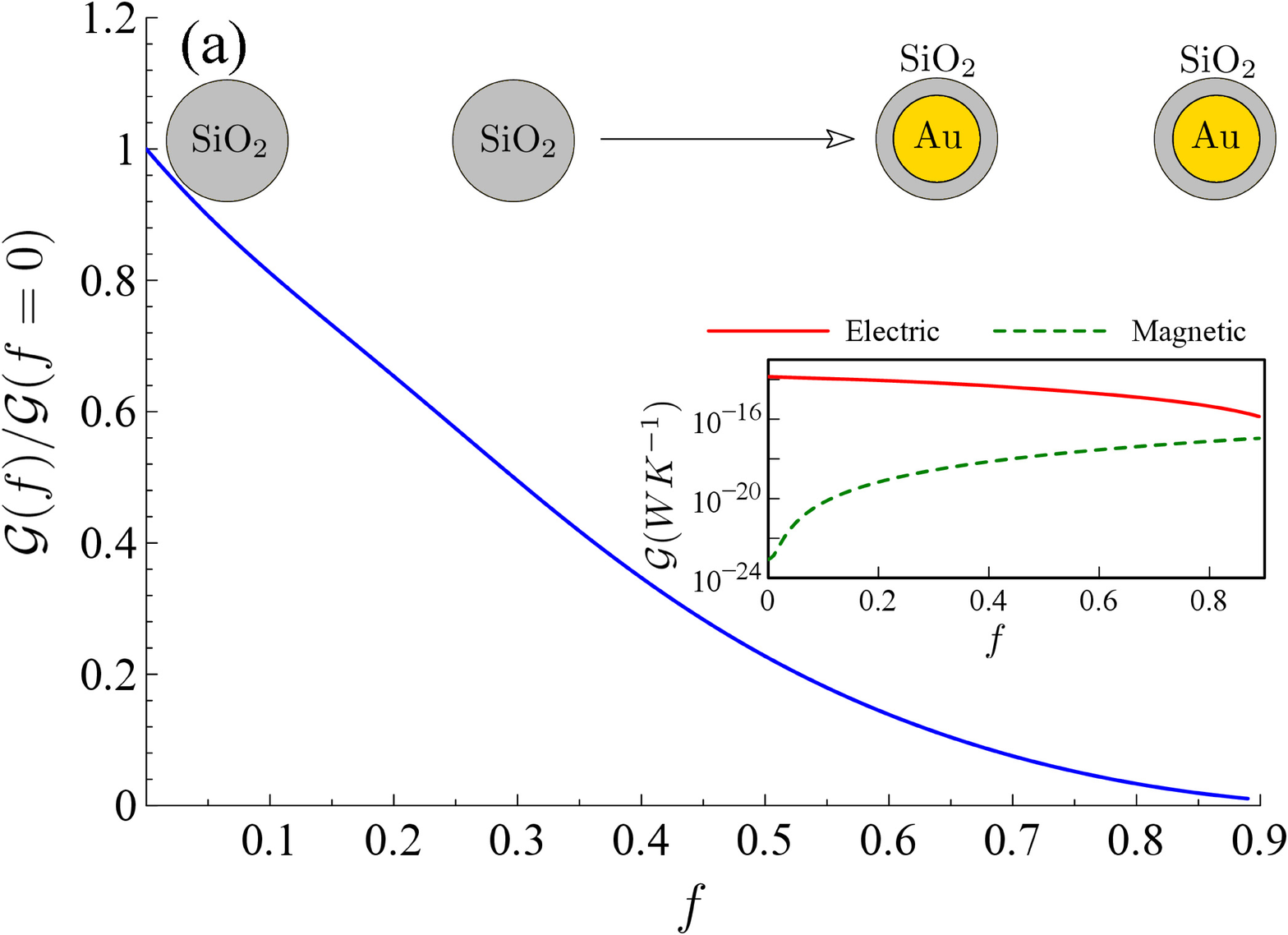}
\includegraphics[height=5.25cm,width=7.35cm]{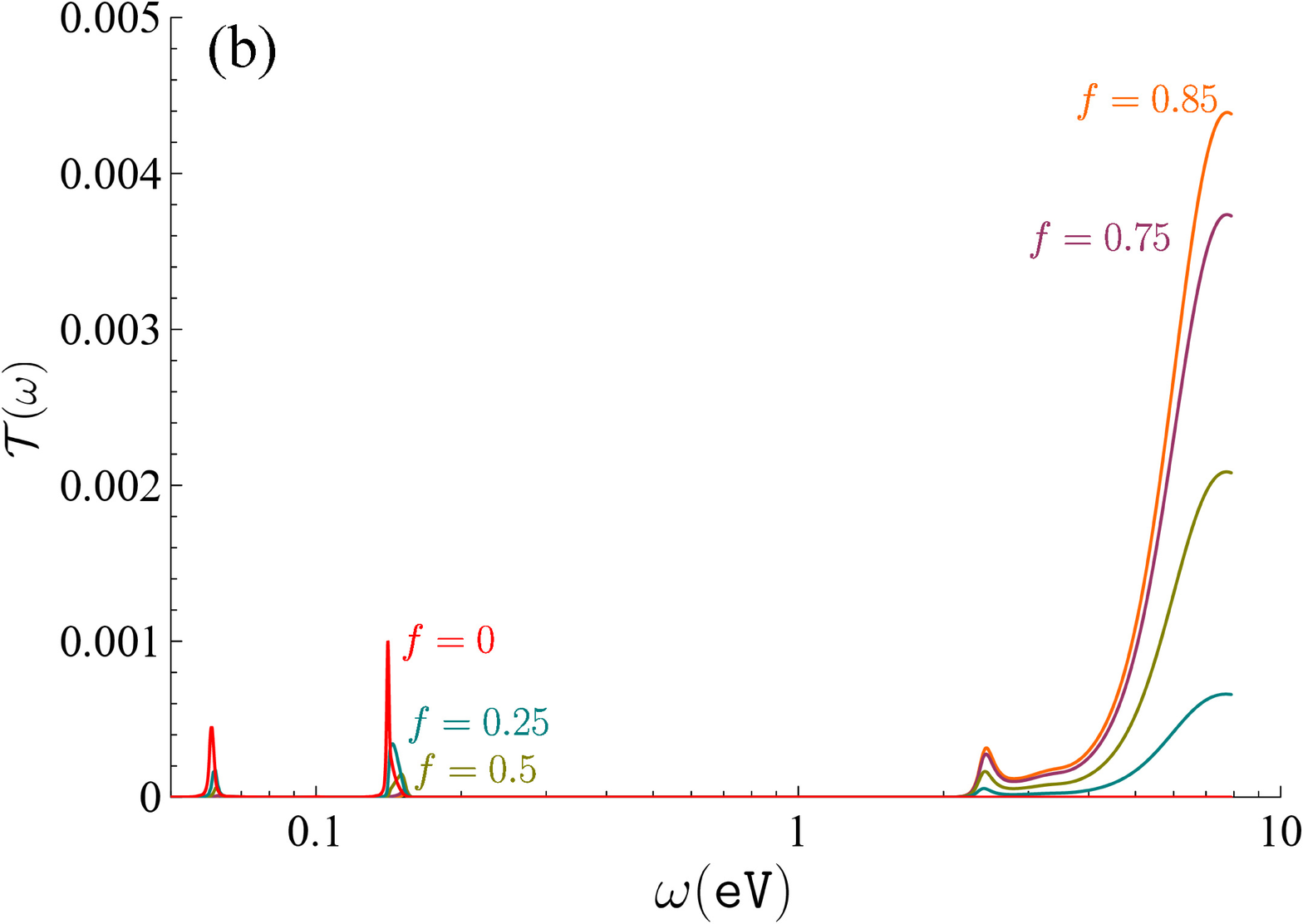}
\includegraphics[height=5.25cm,width=7.35cm]{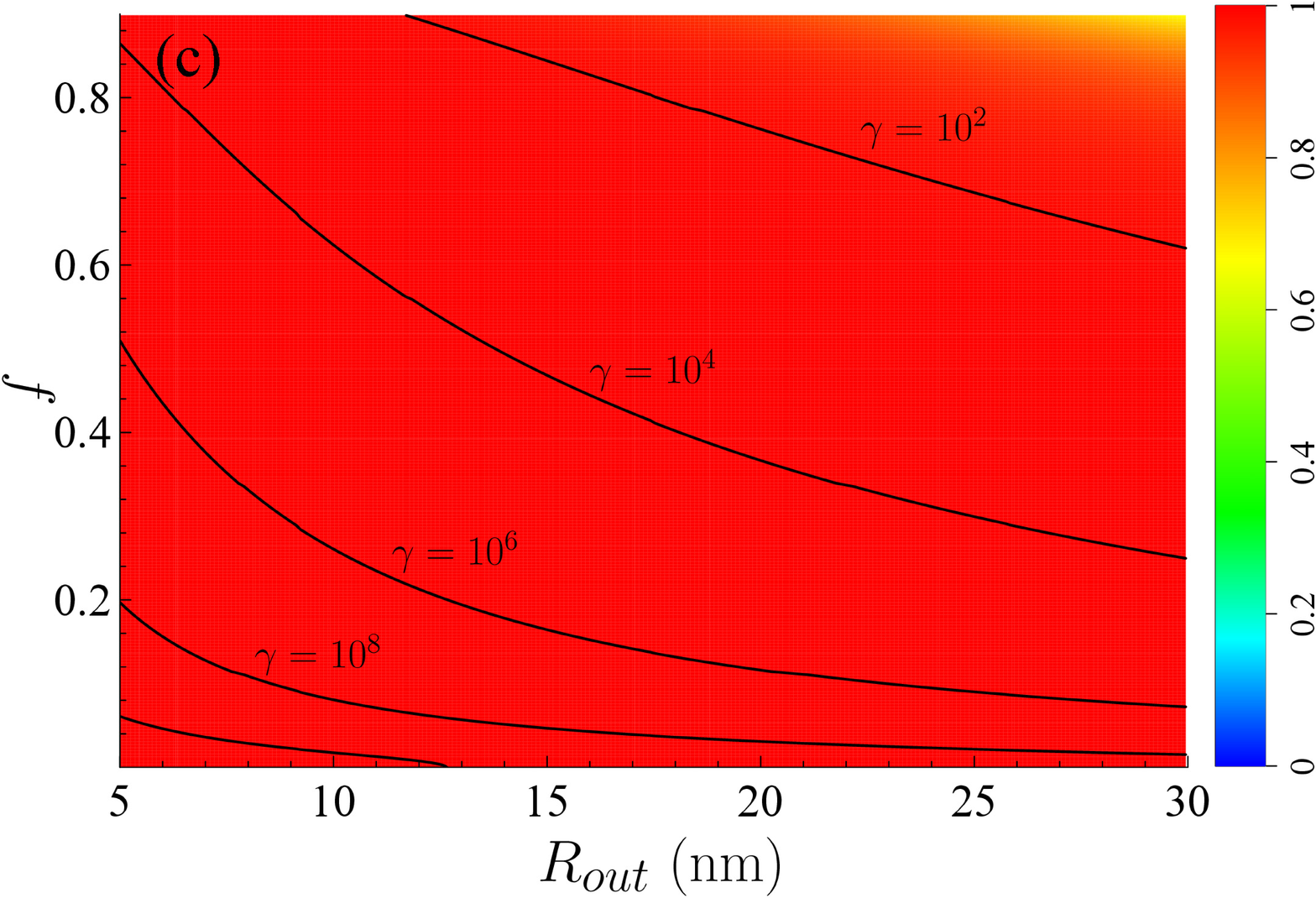}
\caption{(a) Normalized thermal conductance in a dimer of identical Au@SiO$_2$ CSNPs with $d=120$~nm for fixed outer radius ($R_{out}=20$~nm) as a function of gold core volume fraction. The inset graph shows the contribution of the electric ($\mathcal{G}_E$) and magnetic ($\mathcal{G}_H$) dipole moments to the thermal conductance. (b) The transmission probability of the system in part a. (c) Relative contribution of electric dipole to the thermal conductance ($\mathcal{G}_E/(\mathcal{G}_E+\mathcal{G}_H)$) and the ratio of the electric to magnetic dipole contributions ($\gamma=\mathcal{G}_E/\mathcal{G}_H$) in a dimer of identical Au@SiO$_2$ CSNPs as a function of nanoparticles radius and Au volume fraction. }
\label{fig3}
\end{figure}

We first investigate the thermal conductance in a dimer of Au@SiO$_2$ core-shell nanoparticles [see Figs.~(\ref{fig2}) and (\ref{fig3})]. Here, we have taken in account two different cases
to understand the radiative heat transfer in this dimer. In the first case, the inner radiuses (Au core size) are fixed at and the outer radii (SiO$_2$ shell thickness) is varying. The thermal conductance between two spherical gold nanoparticles with radius $R=20$~nm and separation distance $d=120$~nm is shown in Fig.~(\ref{fig2}a) as a function of the silica shell thickness ($t=t_1=t_2$). The result is normalized to the thermal conductance between uncoated gold nanoparticles. It can be seen that covering Au nanoparticles with a SiO$_2$ coating layer results in a strong amplification in the heat transfer between gold spheres. Some more insight
on this amplification is given in the inset of Fig.~(\ref{fig2}a) where the contribution of electric ($\mathcal{G}_E$) and magnetic ($\mathcal{G}_H$) dipoles to the thermal conductance is represented. It is interesting to see that, consistent with the findings of Ref.~\cite{chapuis}, the contribution of magnetic dipole moment is dominant for bare Au nanoparticles. However, by increasing SiO$_2$ coating layer, the electric dipole contribution increases whereas the magnetic contribution does not change significantly. In addition to the dominant and increasing contribution of electric dipole moments, part of the increase in the thermal conductance is due to the increase in the size of CSNPs. As can be seen in Fig.~(\ref{fig2}b), the main reason for the enhancement in the thermal conductance is the increasing contribution of SiO$_2$ resonance modes in the transmission probability between nanoparticles. We also observe that, the surface resonance modes of gold core have slightly red-shifted by silica coating. This is reasonable, due to an increase in the local refractive index surrounding gold nanoparticle produced by the SiO$_2$ layer (in comparison to the host material). However, the contribution of this shift to the increase in the thermal conductance (at $T=300$~K) is negligible. In order to understand the effect of core radius and shell thickness in the heat transfer, we have calculated the electric as well as the magnetic part of the thermal conductance, in the case where the separation distance between
Au@SiO$_2$ CSNPs is $d=120$~nm. In panel~(c) of Fig.~(\ref{fig2}), the relative contribution of electric dipole moment ($\mathcal{G}_E/(\mathcal{G}_E+\mathcal{G}_H)$) is shown as a function of the Au radius ($R_{in}$, horizonal axis) and the SiO$_2$ shell thikness (t, vertical axis). We observe that, the contribution of dipole moment is an increasing function of t (the thickness of silica layer). Moreover, when the silica coating layer has a low thickness ($t\ll R_{in}$), the contribution of magnetic dipole moments is dominant in heat transfer. In the other word, the ratio between electric and magnetic contributions to thermal conductance ($\gamma=\mathcal{G}_E/\mathcal{G}_H$, the black curves in the figure) is decreasing function of the gold size which is in consistent with the findings of Refs.~\cite{heatbetweenmetall2,magneticalphaenhancement,chapuis}.

In the second case, we have discussed the thermal conductance between two Au@SiO$_2$ CSNPs with fixed outer radius ($R_{out}=20$~nm) as shown in
Fig.~(\ref{fig3}a). The thermal conductance is plotted as a function of gold core volume fraction and normalized to the thermal conductance between two SiO$_2$ bare nanoparticles. As we increase the radius of the gold core, there is continuous reduction in the thermal conductance. From the inset of Fig.~(\ref{fig3}a), it is clear that the decrease in the contribution of electric conductance is responsible for this reduction, even though the magnetic contribution (with much less contribution) is an increasing function of gold core volume fraction.
One can see in Fig.~(\ref{fig3}b) that peaks in the transmission probability (due to the resonance of surface mode of silica nanoparticle $\sim 0.2$~eV) slightly blue-shifted and decreased by increasing the gold core volume fraction. Since the plasmon resonance frequency due to the gold core does not changes significantly by size, the reduction in the thermal conductance can be attributed to the decrease in the resonance surface modes of SiO$_2$ layer as it gets thinner.
Results for the electric dipole contributions in heat transfer is presented in Fig.~(\ref{fig3}b), and one can see immediately that the contribution of electric dipoles in heat transfer are dominant at very small Au core volume fraction and can be up to $10$ order of magnitude greater than the magnetic part.

\section{Thermal conductance in a dimer of SiO$_2$@Au core-shell nanoparticles}\label{sec4}

Let us now examine what happens when a homogeneous dielectric core is uniformly coated with a mantle of metallic composition. To this end, we investigate the thermal conductance in a dimer of SiO$_2$@Au CSNPs. The optical properties of these nanoparticles greatly depend on the hybridization of the two surface modes arising, respectively, at the shell-medium (Au-vacuum) and the core-shell (SiO$_2$-Au) interface, for more details see Ref.~\cite{mathematicsoscoreshell}. Due to this hybridization, the plasmon resonance frequency of SiO$_2$@Au CSNPs could be precisely tuned by adjusting the thickness of the Au layer or silica core volume \cite{sio2au3},
which allows one to control the radiative heat transfer.
\begin{figure}[h]
\centering
\includegraphics[height=5.25cm,width=7.75cm]{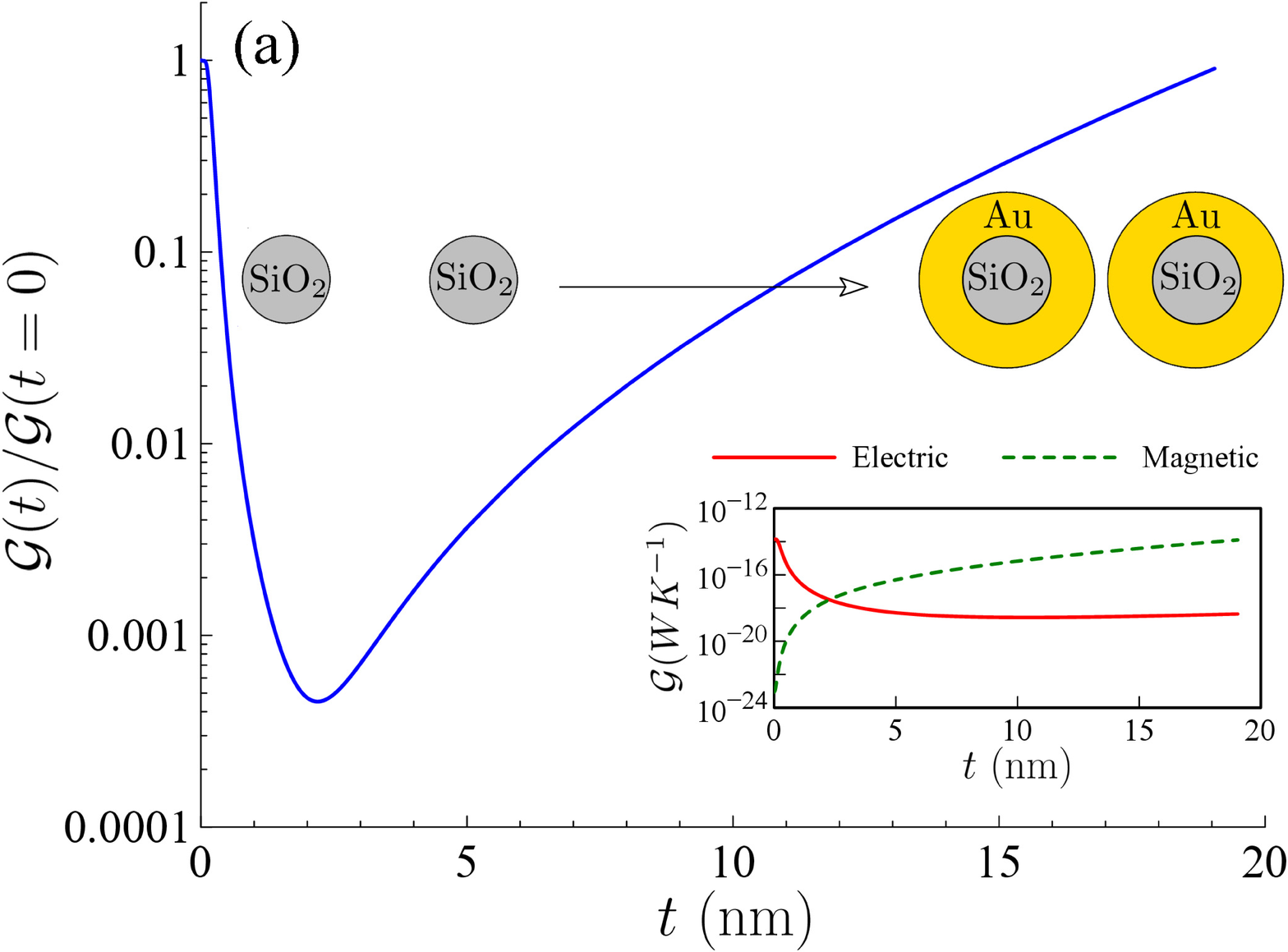}
\includegraphics[height=5.25cm,width=7.35cm]{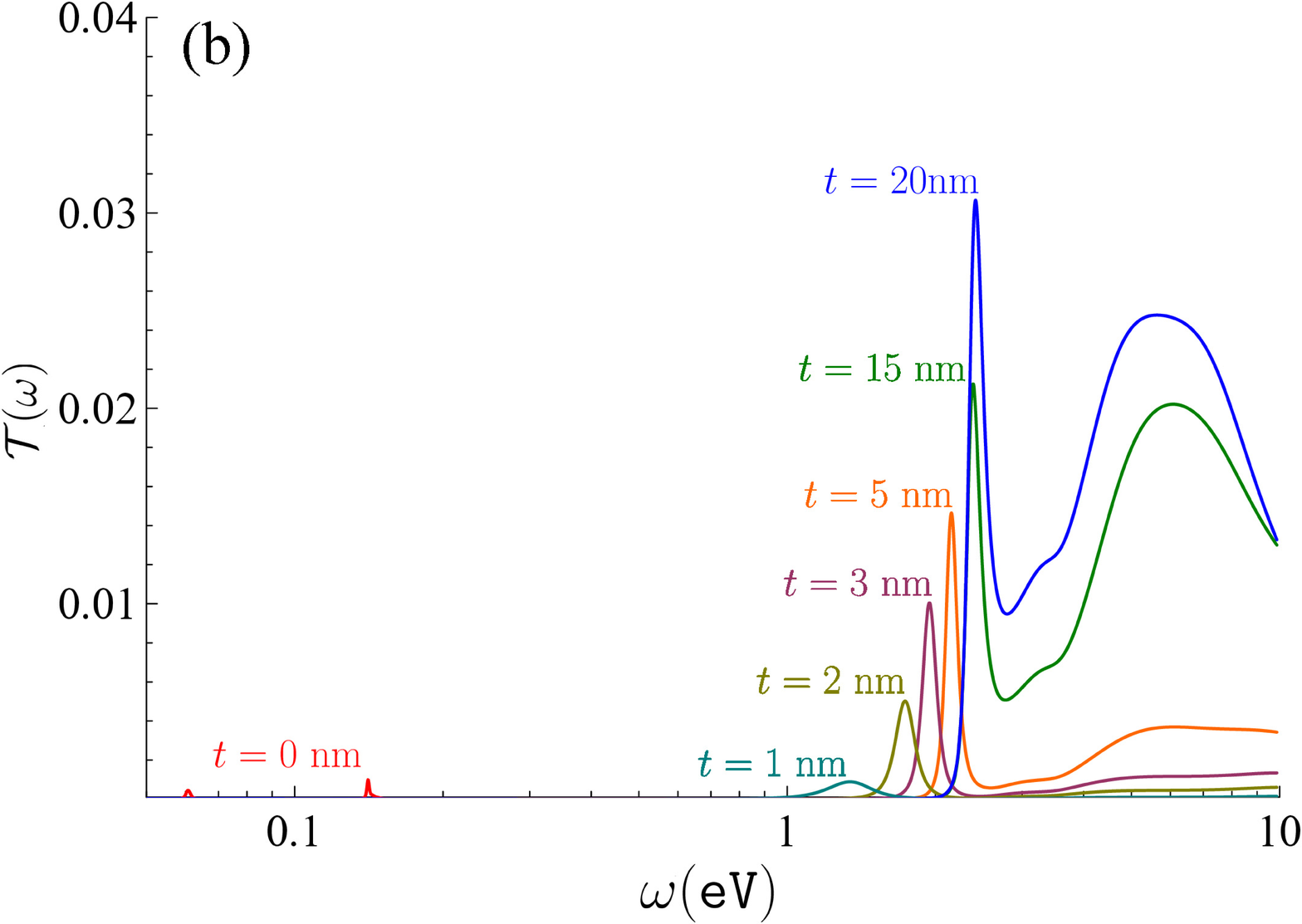}
\includegraphics[height=5.25cm,width=7.35cm]{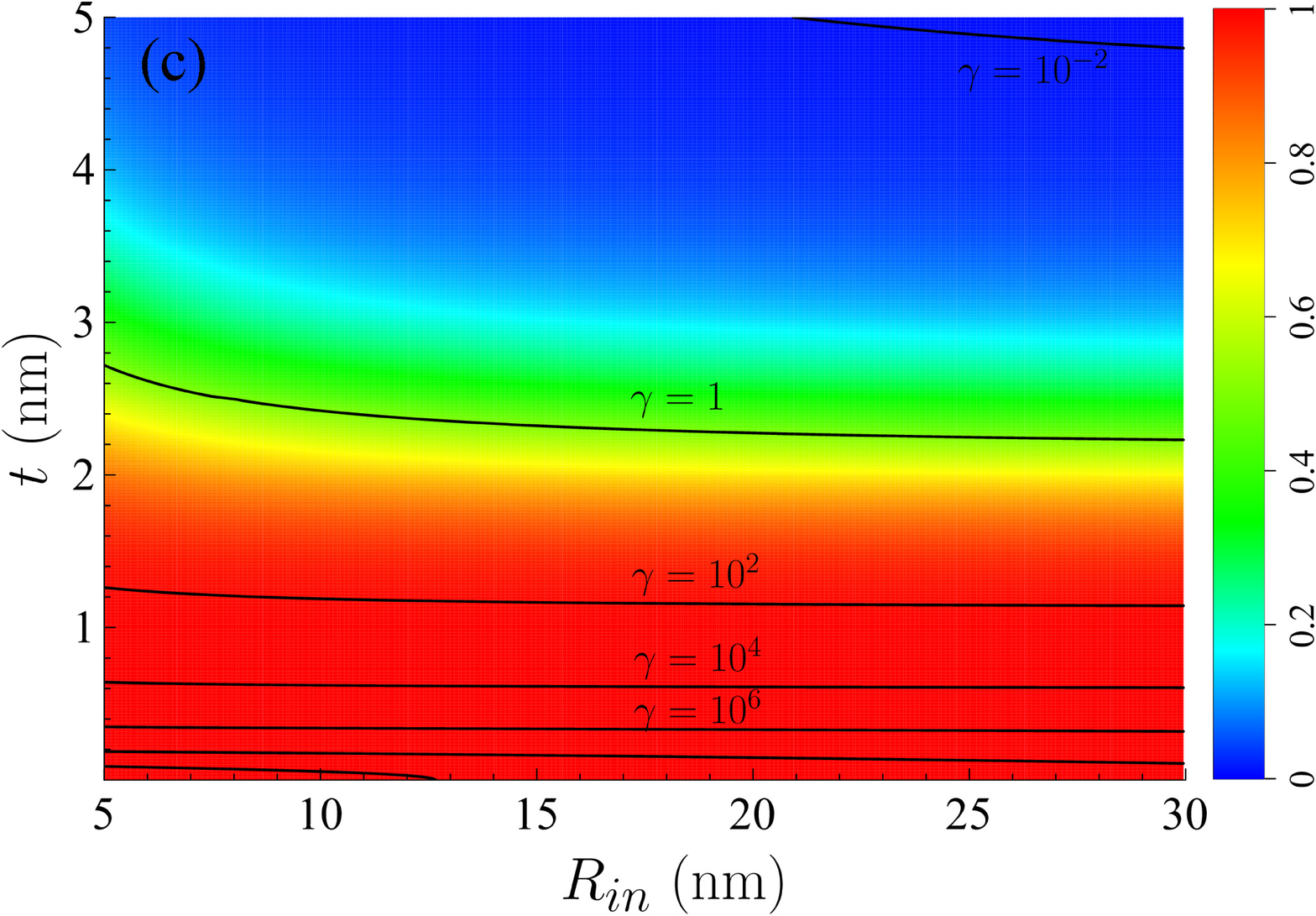}
\caption{(a) Normalized thermal conductance in a dimer of identical SiO$_2$@Au CSNPs with $d=120$~nm for fixed silica radius ($R_{in}=20$~nm) as a function of the shell (gold) thickness. The inset graph shows the contribution of the electric ($\mathcal{G}_E$) and magnetic ($\mathcal{G}_H$) dipole moments to the thermal conductance. (b) The transmission probability of the system in part a. (c) Relative contribution of electric dipole to the thermal conductance ($\mathcal{G}_E/(\mathcal{G}_E+\mathcal{G}_H)$) and the ratio of the electric to magnetic dipole contributions ($\gamma=\mathcal{G}_E/\mathcal{G}_H$) in a dimer of identical SiO$_2$@Au CSNPs as a function of SiO$_2$ radius and Au thickness. }
\label{fig4}
\end{figure}

Figure~(\ref{fig4}a) shows the influence of the gold layer thicknesses on the thermal conductance between two silica nanoparticles with radius $R=20$~nm and separation distance $d=120$~nm. The thermal conductance in a dimer is normalized to that of uncoated SiO$_2$ nanoparticles. Result indicates that coating a silica nanoparticle with a thin gold layer leads to a sharp decline in the thermal conductance, while with further increase of coating thickness the thermal conductance increases. We observe from the inset of Fig.~(\ref{fig4}a), that the thermal conductance due to electric dipoles decreases significantly with increasing gold thickness, and the magnetic dipole moment has a dominant and increasing contribution to the thermal conductance for ($t\gtrsim 3$~nm), which leads to increased heat transfer.

It is clear form Fig.~(\ref{fig4}b), that presence of the thin gold coating layer, shields SiO$_2$ core and yields a strong reduction of SiO$_2$ surface modes to the energy transmission probability. The results are consistent with the reduction in the intensity of dipolar resonance of the bare SiO$_2$ nanoparticles \cite{sio2ausynthesis1,sio2ausynthesis2}.
By increasing the thickness of the gold layer, the SiO$_2$@Au CSNPs exhibit well-known symmetric and anti-symmetric plasmonic resonances, which broaden and display a blue shift with thicker gold shell. The increase in the contribution of symmetric mode in the transmission probability is responsible for the increase in the thermal conductance at larger shell thicknesses. Figure.~(\ref{fig4}c) shows the competition between electric and magnetic dipoles in heat transfer between two SiO$_2$@Au CSNPs. Once again the manifold, $\gamma=1$, in parameter space, separates those parameters for which electric dipoles dominant in thermal conductance and those that magnetic dipoles play crucial role. We can see that the contribution of electric dipoles does not depend much on the core radius, and has a dominant contribution in the heat transfer only in the case of low-thickness shells.

\begin{figure}[t]
\centering
\includegraphics[height=5.25cm,width=7.35cm]{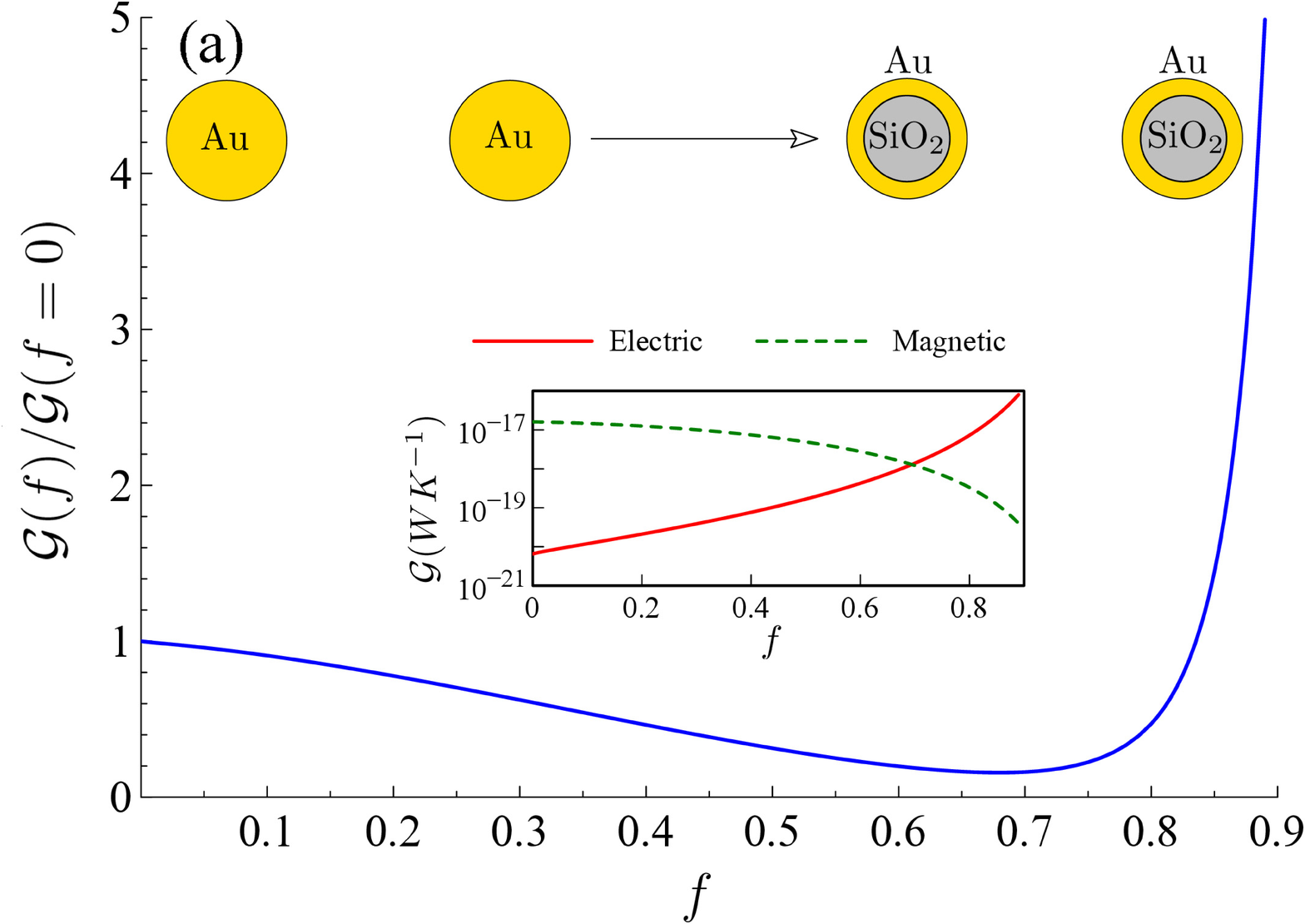}
\includegraphics[height=5.25cm,width=7.35cm]{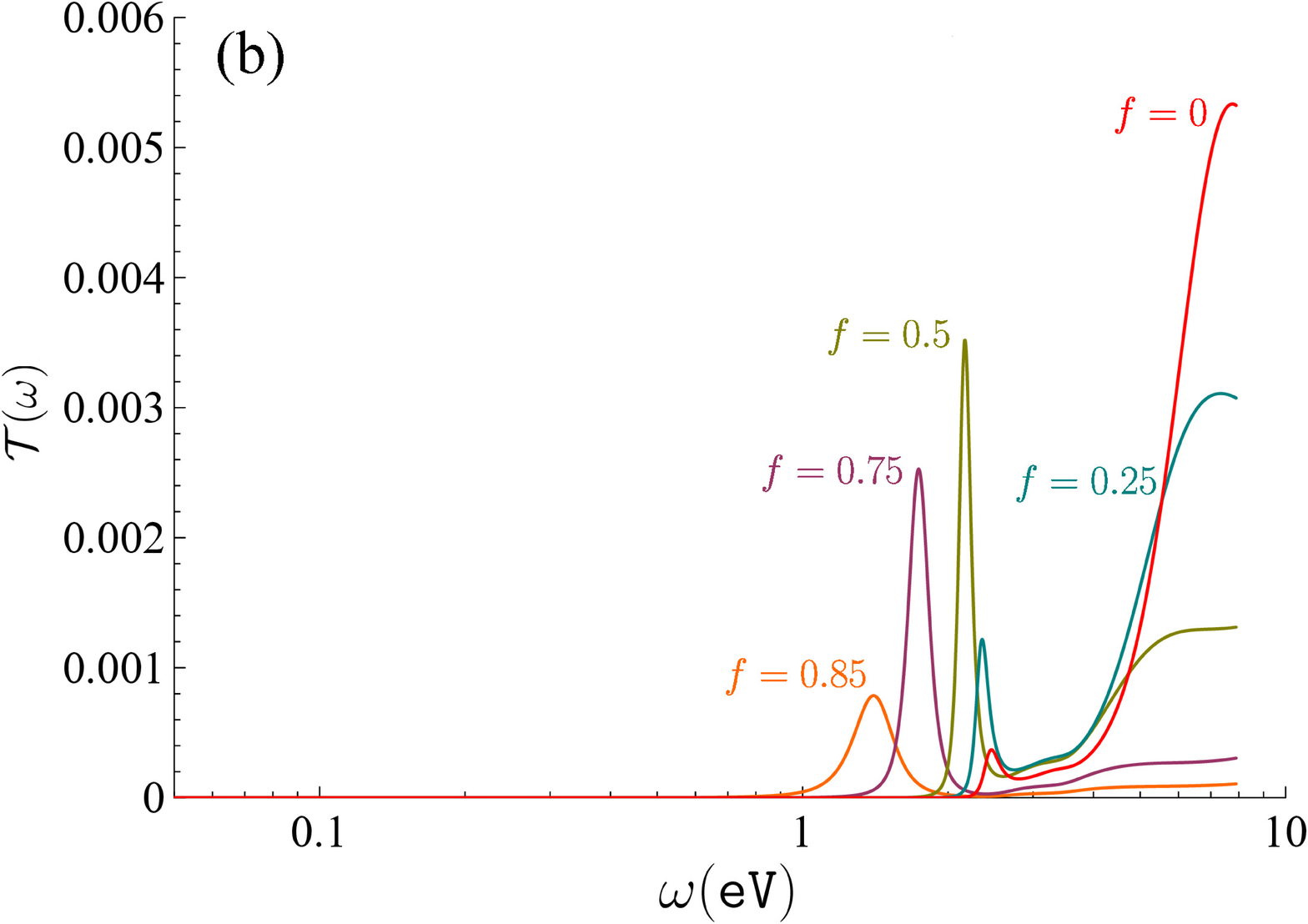}
\includegraphics[height=5.25cm,width=7.35cm]{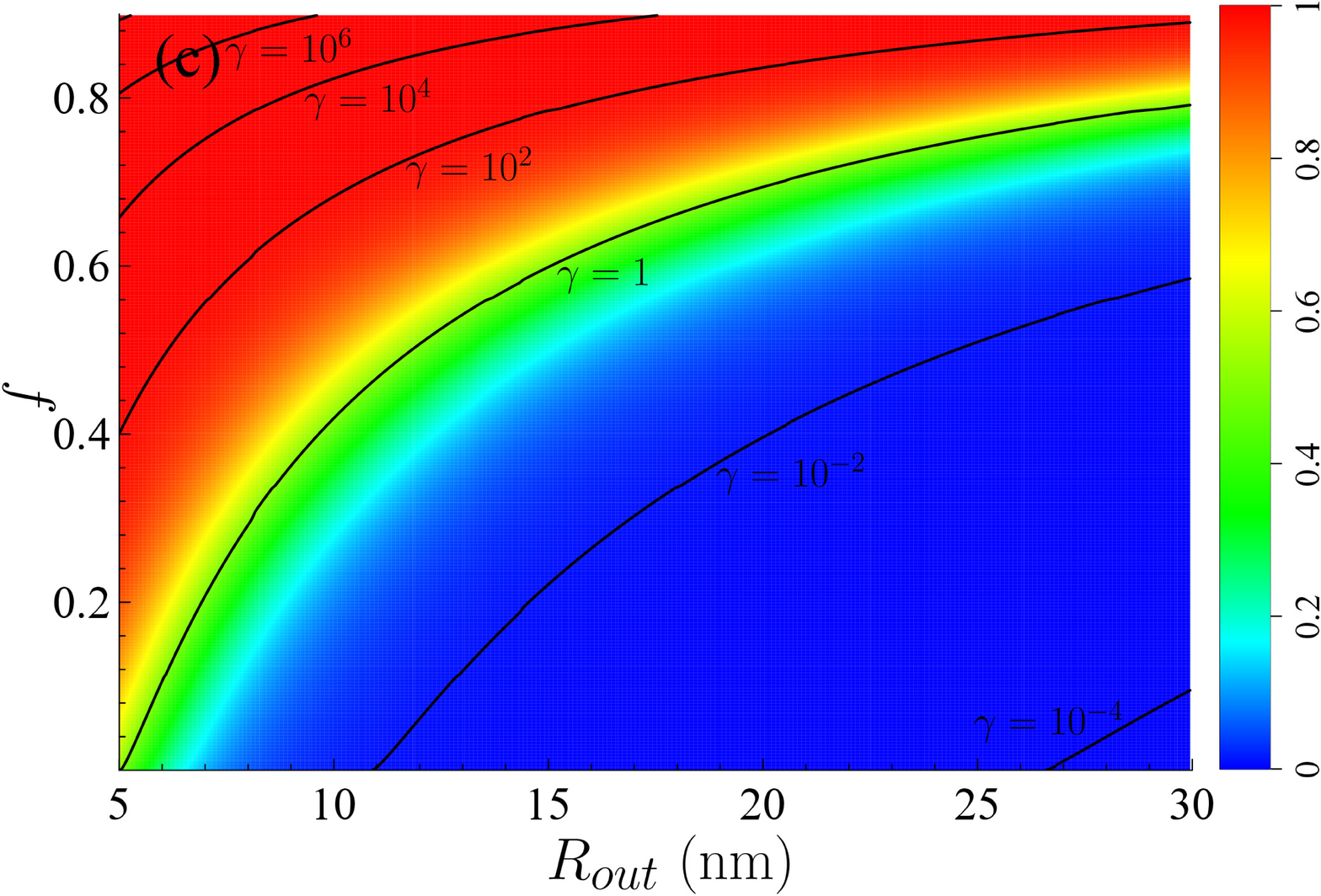}
\caption{(a) Normalized thermal conductance in a dimer of identical SiO$_2$@Au CSNPs with $d=120$~nm for fixed outer radius ($R_{out}=20$~nm) as a function of silica core volume fraction. The inset graph shows the contribution of the electric ($\mathcal{G}_E$) and magnetic ($\mathcal{G}_H$) dipole moments to the thermal conductance. (b) The transmission probability of the system in part a.(c) Relative contribution of electric dipole to the thermal conductance ($\mathcal{G}_E/(\mathcal{G}_E+\mathcal{G}_H)$) and the ratio of the electric to magnetic dipole contributions ($\gamma=\mathcal{G}_E/\mathcal{G}_H$) in a dimer of identical SiO$_2$@Au CSNPs as a function of nanoparticles radius and SiO$_2$ volume fraction. .}
\label{fig5}
\end{figure}
We plot in Fig.~(\ref{fig5}a) the influence of the SiO$_2$ core volume fraction on the thermal conductance between two SiO$_2$@Au CSNPs with outer radius $R_{out}=20$~nm separated by $d=120$~nm apart. For the parameters used here, increasing the core volume fraction up to $f\sim 0.7$ leads to a decrease in the thermal conductance. However, the thermal conductance enhanced for larger core volume fractions. It is clear from the inset of Fig.~(\ref{fig5}a), that the radiative heat transfer due to magnetic dipoles have larger contribution to the total conductance for $f\lesssim 0.7$ [where $\mathcal{G}_E=\mathcal{G}_H$]. For larger SiO$_2$ volume fractions, the increasing contribution of electric dipole moments causes an increase in the thermal conductance. From Fig.~(\ref{fig5}b), it is clear that, transmission probability in a dimer of SiO$_2$@Au CSNPs (with $f\rightarrow 1$) show peaks due to the resonance of hybridized plasmon modes of gold shell, which are red-shifted compared to that of bare gold particles ($f=0$). The enhancement of the thermal conductance at larger core volume fractions could be attributed to the red-shifts in the resonant frequency of symmetric mode of the SiO$_2$@Au CSNP. 

We show in Fig.~(\ref{fig5}c), the relative contribution of electric dipoles to the total thermal conductance, as a function of nanoparticles radius and SiO$_2$ volume fraction. We observe that the relative contribution of electric dipoles is an increasing function of $f$. The relative contribution of electric dipole momemts is decreasing function of the CSNP sizes. Finally, when we look at the ratio of the electric to magnetic dipole contributions, we observe that in the limit of small bare Au nanoparticle, the thermal conductance due electric and magnetic dipole moments have the same order ($\gamma=1$). This result is different from that for Au@SiO$_2$ CSNPs in Fig.~(\ref{fig3}c), where the contribution of electric dipoles is dominant in case of small SiO$_2$ nanoparticles . We find that the contribution of electric dipoles to the thermal conductance increases as $f$ increases. Once again, in the region below the manifold $\gamma=1$, the contribution of magnetic dipoles are dominant, especially for nanoparticles with large radius.


\section{conclusion}\label{sec5}
We have shown that thermal properties of nanoparticles can be further tailored by coating their surfaces with uniform shells to form core-shell nanoparticles. In principle, one can achieve a precise control over the radiative heat transfer between CSNPs by fine tuning the material composition, and size of the core or shell. We have shown that, the contribution of electric and magnetic dipoles to the heat transfer depends not only on the material compositions, but further could be tailored by the core-shell nanoparticle characteristics. It is shown that the presence of the silica coating layer yields a strong amplification for the heat transfer between two gold spheres. On the other side, the thermal conductance in a dimer of SiO$_2$@Au can be reduced compared to that of bare SiO$_2$ nanoparticles. The idea can be used to identify and search for the best possible combination of materials as well as the proper core volume fraction and shell thickness of core-shell nanoparticles for manipulating the radiative heat transfer commensurate with the intended application of CSNPs.

\section*{Acknowledgement}
The author would like to thank Mir-Faez Miri for helpful discussions.


%

\end{document}